# A Dynamic-Growing Fuzzy-Neuro Controller, Application to a 3PSP Parallel Robot

Mohsen Jalaeian F., Mohammad-R. Akbarzadeh-T, Alireza Akbarzadeh, and Mostafa Ghaemi
Center of Excellence on Soft Computing and Intelligent Information processing, Ferdowsi University of Mashhad, Iran
Emails: {m.jalaeian@yahoo.com; akbarzadeh@ieee.org; ali_akbarzadeh_t@yahoo.com}

*Abstract*—To date, various paradigms of soft-Computing have been used to solve many modern problems. Among them, a self organizing combination of fuzzy systems and neural networks can make a powerful decision making system. Here, a Dynamic Growing Fuzzy Neural Controller (DGFNC) is combined with an adaptive strategy and applied to a 3PSP parallel robot position control problem. Specifically, the dynamic growing mechanism is considered in more detail. In contrast to other self-organizing methods, DGFNC adds new rules more conservatively; hence the pruning mechanism is omitted. Instead, the adaptive strategy 'adapts' the control system to parameter variation. Furthermore, a sliding mode-based nonlinear controller ensures system stability. The resulting general control strategy aims to achieve faster response with less computation while maintaining overall stability. Finally, the 3PSP is chosen due to its complex dynamics and the utility of such approaches in modern industrial systems. Several simulations support the merits of the proposed DGFNC strategy as applied to the 3PSP robot.

I. INTRODUCTION

Two most important paradigms in artificial intelligence and soft-computing are the fuzzy inference systems and artificial neural networks [1]. Artificial neural networks are a fruitful approach in approximation, estimation, prediction and most of all modeling in nonlinear dynamic systems [2, 3]. In contrast, fuzzy systems have a special place in handling uncertainties in modeling, decision making and control [4, 5]. Obviously, the synergy of these two powerful paradigms seems to be enthralling; thus two combinations of them have been introduced by scientists, Neuro Fuzzy System (NFS) and Fuzzy Neural Network (FNN), which aim to embrace various merits of both paradigms [6 - 9].

FNN has become one of the most efficient hybrid methods for adaptive control due to its many qualities; namely fuzzy reasoning, neuron-learning and universal approximation [10, 11]. FNN offers a favorable structure for approximation and decision making with finite nodes, when assisted by an appropriate learning algorithm.

In fact, traditional controllers are usually complex in design; and relatively time-consuming because of involving complex mathematical equations. Moreover, if the plant is changed even insignificantly, these controllers must be designed anew. Also the internal obscurity of the plant and external disturbances both lead to an inefficient control, so the adaptive controllers are the only choice to deal with these issues. In recent years, scientists have taken an interest in developing adaptive intelligent controllers; and the FNN structure, as a powerful new adaptive controller, is progressing at a phenomenal speed [12 - 14].

A supervisory controller is using to ensure the stability of the controlled system [12]. As a proper supervisory architecture, Sliding-Mode Controller (SMC) is an effective robust control approach for nonlinear dynamic systems [12, 13, 15, and 16]. The most important merits of SMC are its insensitivity to plant parameters variations, its ability to external-disturbance rejection, fast dynamic response, and design simplicity [15, 17].

Fuzzy neural network structure copes with the chattering phenomenon of SMC and the sliding mode controller guarantees the control system stability using Lyapunov stability theorem [12].

As an important application of the controllers, robot position and speed control is a wide research area including good works and studies in recent years [18]. In 2007, a model based method was introduced which achieved adequate accuracy and efficiency [19]. But such approaches are often too time-consuming in practice; also they are hard to design for sufficiently complex robots. In contrast, model-free controllers were introduced such as in [20]; which could not achieve robustness although they solved the difficulty of the controller design.

After inventing parallel robots by J. P. Merlet [21], due to the complexity of its dynamic equations, parallel robot control has become an active research area [22]. The past controllers must be designed anew for such a fast complex machine. Some studies have been being done in this domain, and the results are improving during these studies; however there are still some perspectives to be desired such as in less computations, faster adaptation, and improved performance.

Here, a novel Dynamic Growing Fuzzy-Neuro Controller (DGFNC) is introduced, as an adaptive intelligent strategy, based on the preliminary work of authors in [23]. More details on the dynamic nature of the strategy are provided, and further simulation studies illustrate the advantages over a special parallel robot control. Finally, stability of the resulting system is guaranteed due to the use of sliding mode supervisory control law.

## II. PROBLEM STATEMENT

Consider the robot equation of motion (1):

$$\tau = M(\Theta)\ddot{\Theta} + V(\Theta,\dot{\Theta}) + G(\Theta) \quad (1)$$

Where $M(\Theta)$ is the mass matrix, $V(\Theta,\dot{\Theta})$ is a vector of centrifugal and Coriolis terms, $G(\Theta)$ is a vector of gravity term and $\tau$ is a vector of actuators' torque. In parallel robots, the equation of motion is more complex than serial manipulators, thereby traditional controller become more complex to design and often are impractical from a real-time computational perspective. Hence, we chose an adaptive intelligent controller.

Position control objective in robots is to move the robot end-effector (tool) towards the desired position with the highest feasible accuracy. The relation between tool and actuators is defined by inverse kinematic equations. Here, the control method is independent joint control, thus we need the inverse kinematic equation for converting desired tool position in Cartesian space to actuators' angles in joint space.

In this study, we apply a novel dynamic Growing Fuzzy Neural Controller to a special parallel robot, 3PSP. The controller is described in part *III* and 3PSP parallel robot is explained in part *IV*, and after that the simulation results is discussed.

## III. THE PROPOSED CONTROLLER

A simple feed-forward fuzzy neural network is used as a controller that has three layers; first layer gets inputs and pass them directly to the next layer, the fuzzy rules effect on the inputs in second layer, and finally, the third layer inferences the controller output by aggregating and defuzzifying the antecedent outputs of the rules. Since membership functions of the fuzzy rules are Gaussian MFs, the output of FNN controller can be shown as Eq. 2.

$$u_{fnn} = \sum_{k=1}^{R} \xi_k \left\{ \prod_{i=1}^{n} \exp\left[-\frac{(z_i - m_{ik})(z_i - m_{ik})}{(\sigma_{ik})^2}\right] \right\} \quad (2)$$

Or it can rewrite as Eq. 3:

$$u_{fnn} = \sum_{k=1}^{R} \xi_k \Gamma_k \quad (3)$$

Where $u_{fnn}$ is FNN-controller output, $z_i$ represents the input of ith node (Error and its time-derivatives), $z_i = e^{(i)}, i = 1,2,\ldots,n$, $\xi_k$ is the consequence weighting of the kth rule, $m_{ik}$ and $\sigma_{ik}$ are the mean and standard deviation of the kth rule belonging to the ith input, $\Gamma_k$ is the output of kth node of second layer, $R$ is the number of fuzzy rules and $n$ is the number of controller inputs (the order of controlled system). So the controller parameter for training contains $\{R, \xi_k, m_{ik}, \sigma_{ik}\}$. Learning mechanism is comprised of two part; Structure learning and parameter learning mechanisms. The aim of structure learning mechanism is to find the optimal FNN structure by determining $R$, the number of hidden layer nodes. On the other hand, the parameter learning mechanism must adjust the parameter considering the determined optimal structure. Learning mechanism is detailed in sections *III.B* and *III.C*.

Since we have to guarantee the controller stability, a simple sliding mode controller has been chosen as a supervisory controller; and the stability is proven using Lyapunov theory (please see *III.A.*).

### A. Supervisory Controller

Consider the following equation as nth order nonlinear system:

$$x^{(n)} = f(X,u) + \Delta f(X,u) + d = \\ \{f(X,u) - hu\} + hu + \delta_1 = f_n(X,u) + hu + \delta_1 \quad (4)$$

In which the state vector of system is $X = [x \; \dot{x} \; \cdots \; x^{(n-1)}]^T \in \Re^n$; $f(X,u)$ is an unknown continues function; $h$ is a given constant and $u \in \Re$ is the input of system. The term of $\delta_1$ comprises the internal system imperfection $\Delta f(X,u)$ and the external disturbance, $d$. If the state trajectory is $x$ and the trajectory command is $x_c$, then we define the tracking error as $e = x_c - x$, and $x_c$ is continues and available. We assume that $\delta_1$ is bounded by $|\delta_1| < D_1$ that $D_1$ is a finite positive constant. A sliding surface is defined as:

$$s = e^{(n-1)} + k_1 e^{(n-2)} + \cdots + k_{n-1} e + k_n \int_0^t e \, d\tau \quad (5)$$

Now for designing a sliding mode controller we have:

$$u_{SMC} = u_E + u_H \quad (6)$$

The equivalent controller and the hitting controller are represented respectively as:

$$u_E = h^{-1}(-f_n(X,u) + x_c^{(n)} + k_1 e^{(n-1)} + \cdots + k_n e) \quad (7)$$

$$u_H = D_1 \text{sgn}(s) \quad (8)$$

Where, $\text{sgn}(.)$ is the sign function. Substituting (5)-(8) in (4) yields:

$$e^n + k_1 e^{n-1} + \cdots + k_{n-1}\dot{e} + k_n e = \dot{s} = -D_1 \text{sgn}(s) - \delta_1 \quad (9)$$

As it was mentioned, control scheme is Independent Joint Control; so three MISO controllers were utilized to control the plant. To check the stability of each controller, we use the Lyapunov function [12]:

$$V_1(s) = \frac{1}{2}s^2 \quad (10)$$

Differentiating the Lyapunov function with respect to time and using (9), we have:

$$\dot{V}_1(s) = s\dot{s} = s(-D_1\text{sgn}(s) - \delta_1) \leq (-D_1 + |\delta_1|)|s| \leq 0 \quad (11)$$

Lyapunov theorem guarantees asymptotic stability. Thus, SMC introduced in (6) can guarantee the stability of the control system. For more details see [12]. If any disturbance is made externally, the SMC controller damps its effects and then the main controller goes on the high-efficiency controlling.

## B. Dynamic Growing Mechanism

The end of structure-learning is finding the optimal nodes quantity ($R^*$) and here it serves this purpose with the Dynamic Growing strategy. Some self-organizing method was introduced by the scientist, as instant, auto-structuring mechanism was introduced by Cheng [12], in which the structure learning is divided to node-adding and node-pruning mechanisms. Although node-pruning mechanism balances the quantity of nodes against to the node-adding, it prunes the trained nodes and this fact makes learning mechanism too time-consuming because each node must train and after some iteration when the error reduced this node will be pruned; so with a little disturbance a new node must be generated and trained again. Proposed Dynamic Growing mechanism gives up the pruning algorithm, and instead it done adding precisely and adapts nodes carefully.

Node-Adding in DGFNC is designed as a function of $R_{cur}$ (Current number of nudes), $\Delta t$ (Run time of each control loop), $Err$ (Error) and $\Gamma_{max}$ (Maximum output of second layer nodes, on other word, maximum output of the rules antecedent).

$$R = f(R_{cur}, \Delta t, Err, \Gamma_{max}) \quad (12)$$

Actually we suppose that R can increase if $C_{add} > C_{th}$; in which $C_{th} \epsilon [0,1]$ is a pre-given real constant and $C_{add}$ is computed as bellow:

$$C_{add} = C_R * C_t * C_e * C_\Gamma \quad (13)$$

$C_R$ is depending on the number of current nodes; it means the generating chance of new nodes become lower as well as the number of current nodes become higher.

$$C_R = 1 - \frac{R}{R_{max}} \quad (14)$$

$C_t$ observes the run time of each control loop; the new nodes must be generated more precisely if the speed of controller is become slowly.

$$C_t = 1 - \frac{\Delta t}{t_{max}} \quad (15)$$

Also, $C_e$ is depending on the error; when the error become to growing, nodes quantity must increase.

$$C_e = \frac{Err}{E_{th}} \quad (16)$$

And finally, $C_\Gamma$ pays attention to the maximum output of rules; if $\Gamma_{max}$ was smaller than a threshold the nodes do not cover all of the input space thus new nodes are needed.

$$C_\Gamma = \begin{cases} 1 & if \ \Gamma_{max} < \Gamma_{th} \\ 0 & if \ \Gamma_{max} \geq \Gamma_{th} \end{cases} \quad (17)$$

Where $R_{max}$, $t_{max}$, $E_{th}$ and $\Gamma_{th}$ are pre-given constants. The parameters of a new generated node is set as $m_{i\hat{R}_a} = z_i$, $\sigma_{i\hat{R}_a} = \sigma_c$, for i=1,2,…,n and $\xi_{\hat{R}_a} = \xi_c$, where $\sigma_c$ is pre-specified constant and $\xi_c = 0$.

## C. Parameter Adaptation

The aim of Parameter-Learning Algorithm is finding the optimal value of FNN's parameters which includes fuzzy system parameters ($m, \delta$) and neural network's parameter ($\xi$) in relation to the optimal determined structure. This task is carried out using back-propagation learning algorithm and gradient-descent method [12, 5], and finally, the adaptation laws are:

$$\dot{\hat{\xi}}_k = \eta_{\hat{\xi}_k} \frac{\partial E_2}{\partial \hat{\xi}_k} = \eta_\xi \frac{\partial E}{\partial y_0^{[3]}} y_k^{[2]} \quad (18)$$

$$\dot{\hat{m}}_{ik} = \eta_m s \hat{\xi}_k \Gamma_k \frac{2(z_i - m_{ik})}{(\sigma_{ik})^2} y_k^{[2]} \quad (19)$$

Where $\eta_m$ and $\eta_{\hat{\xi}_k}$ are positive constants. Based on our extended experience with this robot control simulations, it can be realized that updating $\sigma$ through parameter-learning mechanism has an insignificant effect on improvement of robots' control, and since $\sigma$ value is almost constant, we can omit the $\sigma$ adaptation rule and instead an optimal predefined value can be assigned to $\sigma$ in the process of adding a new node. This makes adaptation mechanism faster.

## D. Dynamic Growing Fuzzy Neural Controller

The main advantages of this hybrid controller include relief from the complexity and hardships of designing the controller, offering a self-organizing controller, an optimal structure that provides optimal calculations, retaining high performance control, overall stability ensured by the supervisory controller-chosen as sliding-mode, and finally being adaptive and model-free thereby eliminating the effect of modeling uncertainties on controller's efficiency [12, 14].

DGFNC consists of two controllers: supervisory controller and FNN-based controller. Sliding mode is chosen as the supervisory controller; and the FNN-based controller as the main controller. The blockdiagram of DGFNC for each actuator is shown in Fig. 1:

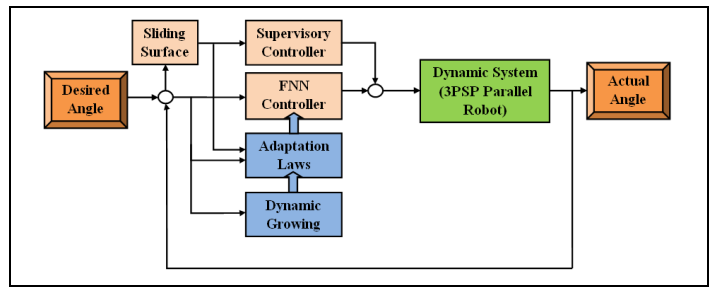

Figure 1. DGFNC close-loop control block diagram.

## IV. 3PSP PARALLEL ROBOT

### A. 3PSP Parallel Robot Structure

Our target plant is a type of parallel manipulator by 3 degrees of freedom named 3-PSP (Prismatic-Spherical-Prismatic). The 3-PSP structure is based on two rigid bodies; a movable platform (star shaped) and a fixed base, which are connected by three PSP legs. In the structure of each leg, there is an actuated prismatic joint and a passive spherical joint paired with a passive prismatic joint. This structure has three closed kinematic loops. For more details about degrees of freedom see [24]. The manipulator has 15 joints; twelve are passive (nine revolute for three spherical and three upper prismatic) and three of them are active and driven by three independent actuators. It has been designed so that the first prismatic joint is actuated in each leg by connecting a motor to a ball screw, (for direct and inverse kinematic of 3-PSP see [24]). Fig. 2 shows the virtual structure of the 3PSP manipulator. This parallel structure has become fully symmetrical. These bars stand at 120 degrees against one another. For defining the position and orientation of the frame, we need to extend or descend each actuated prismatic joint. {T} is attached to the center of the star shape platform by means of a fixed coordination frame and {B} is attached to the fixed base. The base platform was chosen as an equilateral triangle for the sake of simplicity and symmetry of equations which creates a rather symmetrical shape for the 3-PSP robot. Since the manipulator is not to rotate about z-axis of the base frame, the prismatic actuators are welded to the fixed base.

The three degrees of freedom in 3-PSP manipulator are produced by: Changing z-height of point P and also changing the orientation of the moving star about x and y axes of frame {B}.

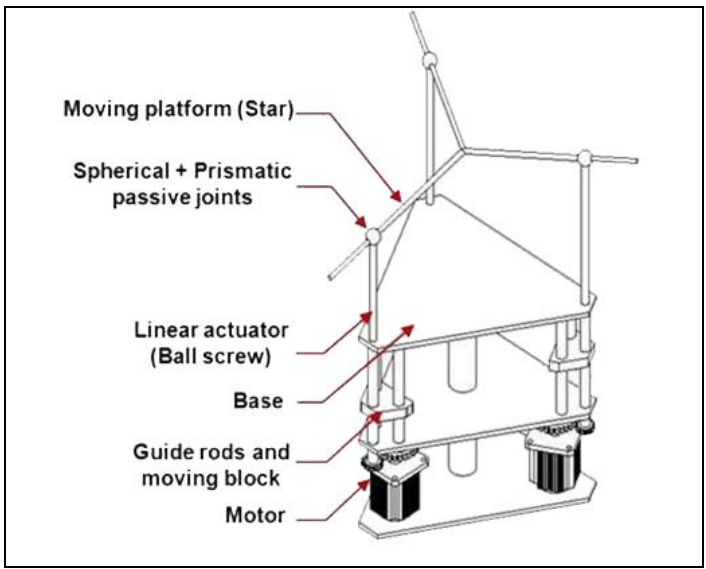

Figure 2. Virtual structure of the special 3-PSP parallel manipulator (This figure is illustrated by Solid Works software based on 3PSP robots model).

The inverse Kinematics objective is obtaining the position of the actuators for a given pose of the end-effector (moving star shape). The control's objective is computing the torque of motors for its determined positions which is obtained by inverse kinematic equations.

### B. Plant Modeling and Simulation

We use the Matlab simulation environment in this study. In a previous study, the dynamic equations were also verified by software such as Solid-works and Adams. Kinematic and dynamic equations of 3PSP are obtained and described [24];

## V. SIMULATION

In this work, we applied DGFNC controller to the 3PSP robot's model using MATLAB software with SIMULINK tools on a computer powered by Intel core2duo 2.2 GHz CPU.

Here DGFNC starts with zero number of nodes in its hidden layer, and pre-given constants are initialized as: h=1, $\sigma_c = 2$, $\eta_m = 0.015$, $\Gamma_{th} = 0.1$, $E_{th} = $1e-5, $C_{th} = 0.02$, $t_{max} = 0.9m\sec$, $R_{max} = 25$.

In this case, the number of nodes grows initially, but reaches a steady state as shown in Fig. 3. Figures 4 and 7 show simulation results for the two Circle and Helix trajectories of the tool position trajectories in 3D space. Also, Figures 5 and 8 show the desired and actual trajectories of joints' angles.

Although there is undershoot at the beginning of joint angle tracking which is zoomed at Fig. 6, this error can be solved by initializing R (number of neurons) at first. This undershoot occurs because actual angles are significantly different from desired angles at the beginning. It is clear that the value of error is more than the amount that DGFNC with very low number of neurons can compensate for. This issue was solved by initializing DGFNC with the certain number of trained neurons, for instance, R initialized by 6 and the $m$, $\sigma$ and $\xi$ matrices initialized by last trained values.

## VI. CONCLUSION

The most important output of this work is to design a self-organizing controller with optimal and swift calculations. We introduced a new intelligent controller, DGFNC, which produced a dynamic mechanism for node-adding. Unlike other self organizing FNN, DGFNC does not lose the trained nodes through pruning-mechanism, but rather improves their learning. Here, the dynamic growing mechanism is discussed in details. The outstanding merits of DGFNC can be described as the optimal and self-organizing structure, brief calculation (a major decrease in time of control loop), as well as proved stability and high-efficiency. Sliding mode approach as a supervisory controller proves stability of the controlled system using Lyapunov stability theorem. Finally, DGFNC is applied to the control of a special fast parallel robot, 3PSP, and simulation results verify the advantages of this new intelligent controller.

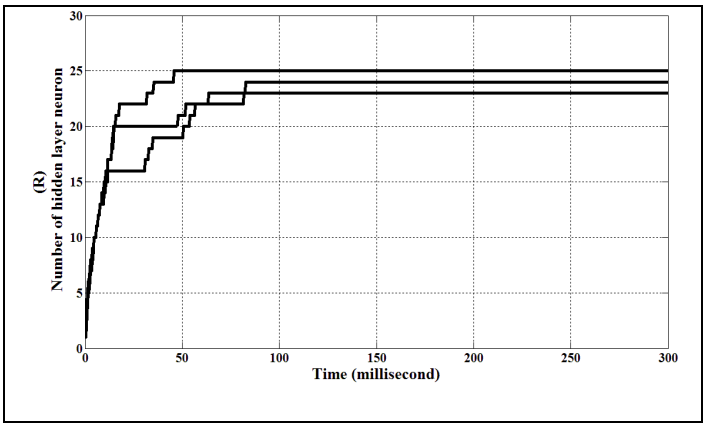

Figure 3. Number of hidden layer neuron through run time.

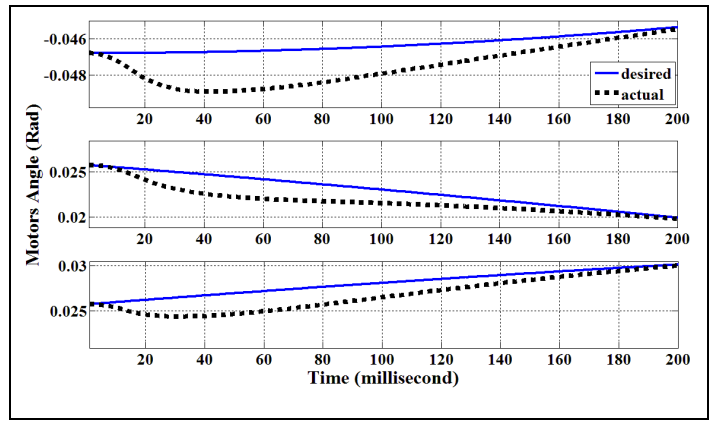

Figure 6. Zoomed desired and actual motors angle.

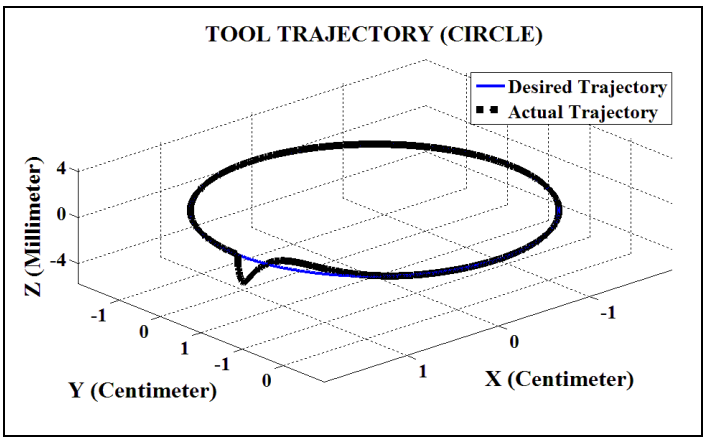

Figure 4. Tool desired and actual trajectories (Circle).

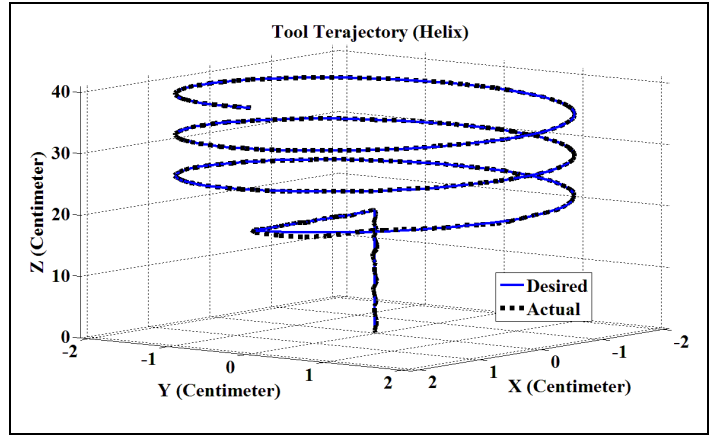

Figure 7. Tool desired and actual trajectories (Helix).

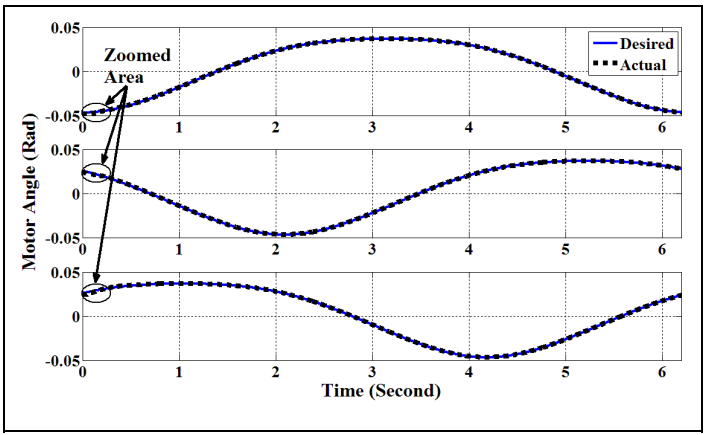

Figure 5. Desired and actual motors angle (Circle trajectory).

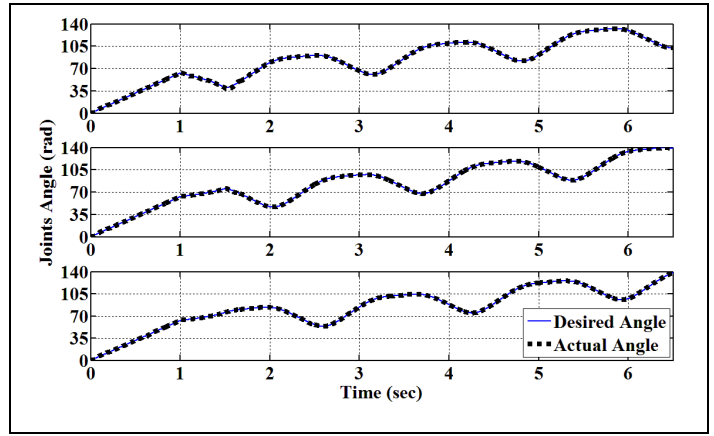

Figure 8. Desired and actual motors angle (Helix trajectory).

ACKNOWLEDGMENT

Last but not least, we would like to gratefully acknowledge the contributions and hard work from our teammates in Parallel Robots Group at Ferdowsi University of Mashhad. This project was made possible thanks to their encouragement. This work makes me realized the value of working together as a team.